\documentclass[aps,prl,twocolumn,superscriptaddress,groupedaddress]{revtex4}
\usepackage{subfig}
\usepackage{graphicx}  
\usepackage{dcolumn}   
\usepackage{bm}        
\usepackage{amssymb}   
\usepackage{slashed}
\usepackage{graphicx}				
\usepackage{amsmath}
\usepackage{mathtools}
\usepackage{tikz,pgf}
\usepackage{comment}
\usepackage{asymptote}
\usetikzlibrary{arrows,backgrounds}
\usetikzlibrary{fit,scopes,calc,matrix,positioning,decorations.pathmorphing}
\usepackage[all]{xy}
\usepackage{yfonts}

\begin{document}
\title{Is there truly classical information?}
\author{Andrei T. Patrascu}
\address{ELI-NP, Horia Hulubei National Institute for R\&D in Physics and Nuclear Engineering, 30 Reactorului St, Bucharest-Magurele, 077125, Romania\\
email: andrei.patrascu.11@alumni.ucl.ac.uk}
\begin{abstract}
All entropy is entanglement entropy. This appears as the result of the existence of black holes. The origin of entropy and the way in which it defines the perceived time direction in macroscopic systems has been discussed and can be debated as long as one ignores black holes. In such a case, thermodynamic entropy may define the arrow of time and entanglement entropy defines some special globally defined entropy encoding information in a non-local sense. With black holes however, if we accept that in-falling objects must have their information encoded outside, for example through non-local fluctuations away from thermalisation of the Hawking radiation, and that the horizon of a black hole cuts all classical information from the outside, then all the entropy that can be added to the black hole, which is basically any form of entropy, must be translated into some form of entanglement entropy to be detected outside. A discussion about the nature of information channels shows that spacetime may not be a correlation-insensitive channel as usually employed in the engineering of quantum computing. Time may emerge as a property of the complexity of entanglement spreading.  
\end{abstract}
\maketitle
The general approach of physics is to look for information about a system within the system itself. It is assumed that by understanding its constituent parts to a higher extent one can know more about the system itself and about physics in general. Therefore, we aimed at higher and higher energy, probing at the same time lower and lower length scales. There are however several results in mathematics that show that information can be encoded not only in the system itself, say, as qubits engraved on some substrate, but also in the context in which the system is evolving, and through all the possible maps that could transform that system into another system considered as benchmark [1], [2]. 
In this article I show that indeed, all information describing physical systems must be quantum in nature and fundamentally global. For that I will describe the emergence of gauge fields as a means for transmitting information across space from a fibre bundle point of view, and I will employ the one known tool that is capable of completely blocking classical information transfer, a black hole. 
The discussion here will be relevant for better understanding mechanisms like, say, purification, with relevance in quantum information [3], [4], [5].
The key example is based on holography and the holographic map [6], [7], [8]. It is well known that if we continue exciting the vacuum state of a quantum field theory, we have no upper bound on the number of particles we can obtain in a limited region of space. However, as we move up the ladder, we obtain enough excitations so that in the end, the entire region will collapse into a black hole. The black hole has some peculiar features. First, it is a thermodynamic system, as shown by Hawking. As such it emits radiation, as a thermal object does. However, the peculiar thing about black holes is that this radiation should in principle be thermal if one follows Hawking's argument. This argument is largely believed to be wrong, although it is not clear in what way it is wrong. If one considers the black hole as a quantum object and one is concerned about the scattering of particles on it, one has to consider that the processes must be fundamentally unitary, i.e. the laws of probability are preserved, or, otherwise stated, there are no probability sinks and sources. If one enforces unitarity, as a guiding principle, one obtains the Page curve, namely the evolution of the entropy of the radiation emitted from the surroundings of the black hole as time goes by does not grow indefinitely. Hawking's calculations show that the Entropy of the emitted radiation should always be thermal in nature and hence grow indefinitely. Black hole unitarity requires that all measurements of the emitted radiation must be consistent with the black hole entropy $S_{BH}\sim \frac{A}{4G}$ [9], [10]. Entropy is a peculiar quantity. First, it is a thermodynamical quantity, and hence it must be associated only with an ensemble, not a single system. By extension we can talk about the entropy of a black hole, but we have to think about it only as the result of the measurement of many systems in an ensemble. Therefore a description of black hole entropy should take into account the black hole microstates. This has been done in the context of string theory [11], [12]. As such, entropy is a good tool for measuring information. This is valid in the context of quantum as well as classical information in an interesting way, when looking at information channels. If we try to transmit information from one system to another, whenever the receiver can fully predict the information received, communication is irrelevant and could as well not occur. Basically, in such a case, the classical entropy would be zero and the receiving system could perfectly well predict the information of the sending system without information being emitted. Naturally, in reality such situations are rare, apparently. There exists another component of the entropy, that has been introduced by means of quantum information, namely the entanglement entropy. This quantity emerges from the analysis of a subsystem of a system that would retrieve information about another part of the system when the two are being separated or analysed separately. Basically we would construct a composite system, and obtain a reduced density matrix of one part of the system by tracing away all the elements of the other part. If the resulting density matrix is non-zero the two subsystems are entangled and they share global information leading to non-zero entanglement entropy. 

Special relativity has been constructed on the idea that the only method of obtaining information about a system is by sending and receiving light signals from it. Both special relativistic information exchange and quantum entanglement are based on some form of interaction transmitting the information across space in a causal manner.  From a quantum gravity point of view we say that the information exchanged through that channel is classical. In special relativity (without quantum mechanics) tracing the one side and focusing on the other would completely separate the first part and retain exclusively the information from the other. In that case the information about the system will be classical in nature and fully separable.
Quantum mechanics allows for a modification of this statement. While information can still only be exchanged via gauge interactions, in a causal manner, quantum entanglement allows for global states of knowledge that cannot be separated once a causal interaction created a correlation (entanglement) between them.

 I will argue here that this is a universal way in which information can be encoded.
 
  Discussions about the entanglement wedge and its approach to solving the black hole paradox can be found in [13], [14], [15]. Here, I will use a rather different vocabulary, although the meaning is not too different.
The fair cases in which information exchange is considered classical (i.e. relativistic causal propagation and quantum entanglement) can only be approximations to how nature really works. 
The fact that this is so is underlined by the fact that black holes exist and that the holographic principle works such that the full information of the in-falling object can be recovered in the radiation around the black hole. In fact even what we consider classical transfer of information by light pulses is a quantum process that must rely on some underlying entanglement. The propagation of interactions themselves must be regarded as a quantum effect in which the dominant contribution is from a pseudo-classical condensation of the quantum information sharing protocols. 
In general quantum gravity effects are considered negligible at our energy scales and at the level of spacetime curvatures found around us. This is only partially true. In fact, there are quantum gravity effects all around us, but we decide most of the time not to think of them in that way, hence we ignore them. One of those quantum gravity effects is entanglement. We prefer to look at it from the perspective of inseparable information linking two subsystems globally, but this inseparability effect has its origin in a geometrical view (and finally a topological view) of both standard model interactions and of gravity. The fact that entanglement cannot be seen completely separated from high energy gravitational effects was postulated in the ER-EPR duality, which asserts that entanglement is equivalent to a wormhole geometry. This statement is rather strong and I know many people still disagree with it. However, let us at least assume that it is valid. In that case, quantum inseparable information is equivalent to a spacetime geometry making use of an extra dimension to link two regions of spacetime. What is this extra dimension? Of course we do not lack extra dimensions, for example in string theory. There is however another construction that adds all the required mathematical structure of extra dimensions: the inner space of gauge symmetries. That space emerges due to a special freedom we have in quantum mechanics, namely to make various choices of the phases of the fields, without affecting their resulting physical properties. By doing so, we basically add a new set of reference frames for this inner space, and we use it to connect regions of our spacetime in a consistent manner, generating via the required gauge connections on the respective fibre bundle, a coupling, be it electromagnetic, nuclear, etc. The field strength of the interactions of the standard model are from a mathematical point of view curvatures. The resulting electromagnetic interaction (hence a light pulse) is basically a connection between fibres on a bundle, that makes the global structure of this inner space manifest, and entanglement is non-separability of sub-systems, no matter what space they are on, inner space, or spacetime. In any case, the fibre bundle which is the fundamental mathematical structure describing all gauge interactions, including gravity, by means of the connections between the fibres, is mathematically speaking, nothing but an obstruction to the separability of the small patches of the space we analyse, when a simple cartesian product is invoked. Basically, the fibre bundle encodes the obstruction to the cartesian separability of large scale structures into smaller pieces (patches). That is also the definition of entanglement. 
Therefore, the mere existence of interactions is the result of the simplest form of entanglement that one can imagine, namely between adjacent space (time) patches. We could see those as "wormholes" because they happen in the inner space produced by the gauge group, and in terms of semantics, that may sound like an "extra dimensional portal", but leaving aside fancy words, it is just our propagation of light pulses and light pulses, as well as any other form of interaction, does indeed entangle various systems. 
However, there is another way to prove the fact that all information is quantum. 
Let us take a classical object, and throw it inside a black hole. The horizon forms a perfect barrier against classical transfer of information. We know that the special relativistic transfer of information outside the black hole is forbidden. That "special relativistic" side of my definition of "classical" is satisfied. But even quantum entanglement is not sufficient (at least in its simplest interpretation) to extract information from the black hole. The in-falling object, due to the extreme expansion of the ingoing spacelike surface passing the horizon (a feature specific to black holes) will never have enough time to interact otherwise than strictly perturbatively, with the Hawking modes that the expanding spacelike surface will bring to existence as pairs produced around the horizon. Those pairs will form the Hawking radiation, but that radiation will remain mostly thermal in nature, not having the possibility to entangle with the in-falling object sufficiently. Therefore even the entanglement "side" of my definition of "classical" is not enough to explain away the Black Hole information paradox. The in-falling object does not perceive the horizon as such, being a global object it cannot be detected by local inference alone. The information associated to the observer will be fully recovered (albeit in a scrambled form) in the outgoing Hawking radiation (slowly in the early stages of the black hole life, quicker in the later stages). However, we just imposed a block to all the classical (special relativistic and quantic) information transferred from the black hole outside, and yet, we could, at least in principle, according to the holographic principle, recover all the information allowing us to reconstruct the in-falling object. Therefore if we filter out all the classical information associated to that object, we remain with all the information about the object. Therefore, quite generally, all information about any object must be fully encoded in a global manner via quantum information mechanisms emerging from quantum gravity. There is no classical (quantum) outside quantum gravitational information. This way of thinking also underlines the quantum nature of gauge interactions. 
On the other side let us see how light propagation is based on the principles of entanglement and why the ER-EPR mechanism is valid even at "room temperature" (aka in the solar system). We remember that the wavefunction or our quantum field had an additional complex phase, specific to quantum mechanics, that allowed us to calculate probabilities involving physically un-realised intermediate states and obtain quantum superpositions that would entail interference patterns, among other phenomena. It is this phase that we use to impose gauge connections on our system, by allowing the wavefunction to be invariant to transformations of those phases induced by certain groups. If we see those groups as generating additional geometric structure, we best describe the transition from one region to the next by means of a fibre bundle with the structure group the gauge group. In any case, entanglement usually involves operations on the same quantum phase, basically first, the application of a Hadamard gate usually generates a superposed quantum state, while the application of a CNOT gate links two different unitary lines, switching the second when the first is in a certain state generates entanglement. A fibre bundle does something very similar: 
It takes two different patches of our manifold, overlaps them in a certain area, generating some form of superposition, then defines a transition function that introduces a global "twist" on the entire manifold, with a certain matching condition on the intersection, mainly introducing the global "twist" when the local intersection matches with the second patch (by "twist" please understand a global "feature" in general, not necessarily a twist, as in the Moebius strip, although that is the most usual example given, knots could be a good example, leading to Jones Polynomials and solutions to Chern Simons theory, not to be discussed here [16]). 
We obtain therefore entanglement in its most obvious form, by simply linking two patches on the gauge bundle by means of fibres and the gauge group. All has been possible because of the existence of the quantum phase in our wavefunction. We can look at the process as entanglement, but we can of course also see it in a geometrical sense, as extending our spacetime with some inner gauge space, performing some local patching that encodes some global structure, and then getting out, at a nearby point. Again, in fancy language we would call that a wormhole, but in reality it is just the propagation of light to an adjacent patch. I hope this makes it clearer how we can understand ER=EPR in every day phenomena, including interaction propagation. This also means that quite generally, spacetime is a network of entangled patches. If a spacetime black hole were to form, we would obtain a spacetime horizon that would separate the regions in spacetime. That means we would separate the first order propagation of light pulses from one patch to the adjacent patch. This is a quantum process, and it is of course causal in the sense of special relativity. The propagation of light defines the special relativistic light cone and the causal structure, while quantum entanglement correlates objects that may be otherwise separated but have been in causal contact at some point. This is what I call "classical" and in some context "first order quantum gravitational" effects. Usually some misunderstandings occur here and one prefers not to talk about topology changes in black hole formation. That is mainly because at the horizon, light can get in, but cannot get out, or at least not in a classical sense. However, mathematically the definition of what we call "open sets" can be made such that it implies "connection by light in both ways". If that definition of open sets is used, then of course the formation of a horizon generates a change in topology, only not in the usual topology commonly understood when discussing spacetime. However, a topology is defined by its open sets, and therefore that is as valid a topology as any other. The transition functions of the fibre bundle obey the cohomology condition, linking at least three patches. At this level, the information is neither lost nor blocked, but continues to be shared in the same way. This would be a higher form of entanglement, hence a stronger correlation than the type that generates light propagation across adjacent patches. I would be curious what would happen if we created such barriers at the next order, in an even more global manner. From this point of view "classical" communication by means of light pulses represents just the patching of regions of the base manifold by means of connections on the fibre. Entanglement means just a higher order form of patching, in which the global structure is involved at a higher level, including for example, cohomology conditions on the transition functions, etc. A spacetime "barrier" like a black hole horizon is no reason to block such global information from emerging, and in fact, it does emerge. The holographic principle stating that we can recover the full information of the in-falling object in terms of a quantum field theory outside the black hole, together with its quantum correlations, basically just means that in essence all information in the object is quantum, the series development just needs a re-summation by the black hole into one that reflects more the "global" structure, to overcome a spacetime barrier set up by the horizon. Unfortunately the mechanisms by which a black hole does that also scramble the information quite badly. 

To make it clearer, in a fibre bundle we have transition functions. Two patches of spacetime become superposed and entangled by the transition functions which creates light propagation and the usual entanglement (emerging from causal connections, say by light pulses). Those transition functions have to obey a series of higher order relations, usually determined by higher degree cohomology relations. One of those will link three patches instead of two, and will have to satisfy a set of consistency conditions. Through these conditions we introduce an even more global structure to our information, one that is not simply recovered by the usual non-separability condition of "classical" entanglement. That type of correlation is even less bound to the causal structure of the light-cone. 
Mathematically, this relies on the definition of transition functions in fibre bundles: let X and F be topological spaces and $G$ our (gauge) group that can act on our fibre space F. We can define an open cover $\{U_{i}\}$ on our initial space $X$ and the transition functions that connect the intersection (superposition) of the two patches and produce values in the group $G$
\begin{equation}
t_{ij}:U_{i}\cap U_{j}\rightarrow G
\end{equation}
Now, the transition functions need to obey a cocycle condition 
\begin{equation}
t_{ik}(x)=t_{ij}(x)t_{jk}(x)
\end{equation}
where $x\in U_{i}\cap U_{j} \cap U_{k}$. The transition functions however can carry more global information than the inseparable information of "classical" entanglement. This can essentially induce a tower of complicated topological structure that has various forms of inseparability, some of it non-detectable by simple quantum mechanical entanglement, say for example that of Alexander's horned spheres, a situation in which a tower of higher order non-separability would emerge. The connectedness of such a structure would indeed be quite tricky. While the union of the horned sphere and its inside is a simply connected object (any loop can be shrunk to a point and remain inside, or, in a quantum information prescription, we have fully separable states, the exterior would not be simply connected and hence a loop outside the sphere would not be reduced to a point outside the horned sphere, and hence would present what we would call entanglement. Yet, even the interior would encode global information. This would generate a tower of "channels" of information transmission other than via a causal connection (and entanglement) that has not been considered before. 

In what follows I will explain the mechanisms and techniques that lead to the same conclusions as stated above. 
It is worth mentioning that in this article I do not offer a solution of the Hawking information paradox. There are good intuitions regarding the fact that the evolution of Hawking radiation must be unitary and entanglement is certainly involved in information retrieval, but that doesn't mean the paradox is solved in any methodical way. What I do here is to use the most likely outcome of the information paradox, namely that unitarity will ultimately prevail and that information can be restored, as well as the holographic principle, to derive an intriguing result about signalling and interaction propagation, namely proving that those must be fundamentally quantum processes. 
Such processes may however, in a quantum gravitational context, involve forms of entanglement (non-separability of some form of global information) that are not mappable or analysable in terms of what I would call "classical" information transmission, via light channels and using the usual entanglement seen in basic quantum mechanics. 
I also show that entropy, understood also as informational entropy, is ultimately entanglement entropy. This has also an impact on the nature of purification procedures in quantum information. 
First, it is worth noting how the radiation around the black hole emerges. The process is more assimilable to the Schwinger process of pair creations, each pair being created from the vacuum, and therefore being entangled with its counter-part. This will generate a wavefunction of a very special type, that will encode information over a large region of space around the black hole, always pushing the created entangled quanta away from the matter forming the black hole and from the previously created pairs. This means entanglement around the black hole continues to increase as new pairs are being created (which is part of the paradox). This already differs radically from any thermal radiation emitted by a normal black body. 
To understand this it may be useful to construct a model for the evolution of matter around a black hole. For this, let us consider a spacelike slice outside the horizon of the black hole. This will be characterised by a fixed time $t=const$ and will extend in space towards the horizon. The black hole has a peculiar property, that cannot be recovered in Minkowski spacetime. Namely, when passing the event horizon, the only possible direction is towards the centre of the black hole (although not necessarily radially so). A similar situation happens outside the horizon with the time direction. We are also forced to move in one overall sense in time, although we can of course slow that movement down by means of Lorentz transforms. In the same way in which the time evolution of a state in Minkowski spacetime is from a past towards a future, the evolution of an object inside a black hole is from the event horizon towards its centre. Therefore, the timelike and spacelike directions are changed. Once crossing the horizon, our spacelike surface is characterised not by $t=const$ anymore, but by $r=const$ while the evolution from one spacelike surface to the next will bring the system inevitably closer to the centre of the black hole. To understand what happens in a black hole one has to consider a way in which those two surfaces can be connected across the horizon. This can be done in a region around the horizon by a surface $\mathcal{C}$ that connects the two slices. However, if we consider two adjacent slices, separated for example by a distance $\delta$ in the region inside the horizon and by a difference $\delta t$ in time in the region outside the horizon, the evolution of the surface from $x\rightarrow x + \delta$ and $t\rightarrow t+\delta t$ will result in a stretching of the region $\mathcal{C}$ in a spacelike manner. We can of course imagine Fourier modes of free fields on those surfaces, of wavelength $\lambda\sim l_{Pl}$ close to the Planck length, or in any sense, small enough to not be excited out of the vacuum, will become also stretched to wavelengths that allow them to become real particles produced by the stretching of our spacelike vacuum surface. In this case, we obtain a pair of entangled particles, that can be considered maximally entangled, and if the stretching continued, those particles will be pushed outwards and another entangled pair will be created. There are two aspects of major importance here: first, the in-falling matter creating the black hole is far away from the region where the pairs are being created, and second, the stretching pushes the entangled pair apart when the next pair is being created. Therefore, black holes do almost everything to make causality hard to recover. Basically, we can assume that the information carried by the entangled pairs emitted by the connecting region $\mathcal{C}$ represents the information of the in-falling matter, however that information is being created in a region that is out of reach to the in-falling matter that most likely already reached a region very close to the interior singularity. Even if by some local perturbations we could create some form of entanglement between the in-falling matter and the created particle pair, that entanglement must have been extremely weak (aka perturbative) in nature, if it was generated by interaction propagating (like an exchange of photons, or gravitons, or whatever) at the speed of light. This must be so simply because the space between the in-falling matter and the particle creation region is expanded by the black hole hence any "first order" propagation of information will have a hard time catching up with it. Second, if the full information is to be recovered in order to preserve unitarity, the whole information must be encoded in entangled pairs created by the vacuum, in a way that is causally disconnected from the in-falling matter. This leads to the old dilemmas of Hawking's radiation. What happens when the black hole reaches a maximal evaporation? One argument claims the existence of remnants, some interesting objects that must preserve some form of entanglement with everything going out in the radiation emitted by the black hole throughout its history. This involves a rather large entanglement entropy, basically, unbounded as one can throw any quantity of matter in the black hole. However, such a remnant has a finite and finally bound spacial extension and hence wouldn't allow for all those inner degrees of freedom, in any approximation. One possible explanation of this phenomenon may be provided by the so called Wheeler's Bags of Gold, which basically just "make room" for more degrees of freedom in the remnant. On the other side, if we assume that the black hole will radiate away all its radiation, we will have quite a lot of outgoing Hawking radiation entangled with basically nothing. The problem here lies in the fact that we could have sent a pure state initially, but after all the radiation is emitted we obtain a mixed state, by a process that we expected to be unitary, which leads to unitarity violation. 
This could probably give a basic idea of the so called information paradox. In essence we do expect for the evolution to be unitary but it is also hard to expect remnants as they do not truly behave like normal physics would demand them to. Remnants create various problems as they would contribute to additional divergences in loop calculations for quantum field theories, etc. 
Black holes make us to some extent more comfortable with a series of phenomena that are at odds with normal physical thinking, among those, the concept of locality. Non-local effects are nowadays found in string theory and in black hole physics, albeit usually they are exponentially suppressed. 
In any case, there is one aspect that this problem also creates: albeit highly scrambled, the full information regarding any in-falling object can be recovered in entanglement patterns of the out-going radiation, produced outside the reach of the in-falling object. This makes us think that in any other circumstance, even in the absence of a black hole, the full information of a system must be completely encoded in terms of entanglement and quantum protocols but also that such quantum protocols must have a stronger global component than what basic quantum mechanics would allow. As we can recover all the information from in-falling matter, and there is technically no restriction as to what information can fall in a black hole, all information must be resulting from some form of entanglement. This conclusion sounds paradoxical in various ways, mainly because we know that we need to use at least some classical information sharing to produce an entangled state and that in general, quantum communication protocols are based on sharing of at least some form of classical information (which is why I call here quantum correlations commonly used in quantum mechanics protocols "classical" and I ask for pardon for possible misunderstandings). It seems that if quantum vacuum and stretching spacelike surfaces are being used, the amount of classical information required for the transmission of a quantum message shrinks to zero. This is not what we usually see in quantum communication protocols, yet, it seems to be what a black hole routinely does. 
Indeed, if we regard gauge connections and their resulting gauge fields to be fundamentally defined through entanglement, hence an obstruction to the cartesian pairing of two independent base manifold patches, then we can agree with the fact that even what we would call classical information transmission, is basically quantum in some sense. If however there is the potential of full quantum entanglement communication, as presented by a black hole vacuum, where no or perturbatively little classical information is exchanged, then what makes the rest of the world be bound to an entanglement that so well simulates a classical relativistic causal structure in which entanglement is included?
I do not pretend to have an answer to this question. Let us understand the role of entropy in transmitting a message. In classical communication, a message is transmitted as a series of bits of different values (from the set $\{0,1\}$ to preserve the usual conventions). The entropy per bit in such a situation will be 
\begin{equation}
S=-\sum_{i=1}^{2}p_{i}log(p_{i})
\end{equation}
where $p_{i}$ represents the probability for a bit to be observed in a certain value. If the message contains $N$ such bits one can extract a total of $N\cdot S$ bits from such a message. In information theory, also, the information sent from the emitter to the receiver is funnily enough associated with randomness. That is because we assume an "innocent" channel that has no way to interpret the information sent, and will assume that information is anything that is not too ordered. An ordered, say, unimodal signal, like a sine wave, indeed cannot transmit any information, hence, what the channel should do is to transmit the "noise". Information entropy therefore becomes a measure for the meaningful information that can be sent through a channel (meaningful in the sense of "random" or unpredictable). But information is never random noise. The whole field of data science is usually concerned with extracting the meaningful information from what we consider random noise, which is by itself not an easy task. In fundamental terms however, we are more than happy to transmit everything that is not perfectly ordered, and hence perfectly predictable, as "information", and to describe it via the entropy of the message. This is perfectly natural from an engineering point of view, but if we consider a more fundamental approach, we note that we basically never have pure randomness (at least in the context of classical information transfer). In fact, analogical generation of pure random numbers is a complicated task (of course, the complications depend on how precisely one wants to obtain a random signal). Anyhow, most of the time, in nature there exist correlations between the data sent. Language is not a random set of bits, for example, but it relies on the contextual understanding of its correlations by the parties involved. What if a special type of channel would be created that would be by construction made to be sensitive to the correlations of english language? Communication through such a channel would be far more efficient, and in essence it is already being done in the field of automated translation, by using neural networks. Given such correlation, what would be the informational entropy? Calculating it would not be an easy task, particularly because there are two effects contributing. The channel being sensitive to certain correlations would decrease the classical entropy, apparently reducing the information that could be transmitted, but that information would already be encoded in our channel, so, there would be no need for it to be transmitted further on. The other factor is the two parties involved in the communication process. They will hold some information, not only about each other, that would facilitate the communication, but about the context of their communication and the over-arching global data they would share. In thermodynamics, entropy represents the obstruction of determining one micro-state if a macro-state is fully characterised. This definition is consistent with the idea of information, as fully determining a state, macro- or -micro, basically means decreasing the entropy and having no "surprise" information that could be shared. The fact that the channel, if assumed to not be fully unaware of the correlations transmitted through itself, could play an important role was considered even by Shannon himself [17]. I introduced a somewhat different notion of entropy, one in which topological information is undetermined and hence can act as "surprise information" in certain contexts [18]. As is probably well known to topologists, various invariants can detect certain topological structures while hiding others. So is the coefficient structure in co-homology, which can reveal certain global properties of a system and hide others. Such an entropy, constructed based on the ability of determining the global properties of a state would clearly be useful in describing information transfer through a channel that is sensitive to relevant correlations in the data. Let us however return to our problem: matter, with all the information it retains, falls in a black hole and is pulled towards the centre such that its distance to the region around the horizon is rapidly increasing. Information regarding that same matter however emerges in the form of entangled particles from the vacuum in the spacelike sheet stretching region. Those particles will ultimately encode all the information about the in-falling body and release it outside the black hole, in a manner that is clearly not causally permitted. What we assumed however, was that the channel of communication was completely "unaware" of the type of information sent and of the correlations it contains. Could that truly be so? The fact that the information about the in-falling objects is scrambled but fully recovered outside, strongly suggests that whatever we considered classical communication regarding the object thrown inside the black hole, is basically equivalent to quantum information obtained via entanglement, outside, with no aid from classical communication, but potentially with some form of channel that is sensitive to the correlations sent through it, correlations that may emerge from higher topological constructions. I strongly suspect that the most natural channel would be such a channel that is already sensitive to the types of correlations transmitted through it, up to the point that there is no need for classical information to be exchanged in extreme situations. This channel could be spacetime itself. Let us consider spacetime as a network of entangled regions, so that gauge connections can be formed, in the geometric understanding offered by a fibre bundle. The gauge connections that can be formed are not random. There is a whole field of research looking into gauge symmetries, the way they can be fixed, and how the parameters of those symmetries behave [19]. The literature in the field is vast, including my work [20]. We have for example un-free gauge symmetries, in which the action of such a symmetry on the action functional only vanishes in the case in which the gauge parameters obey certain differential equations. We can link those unfree gauge symmetries with symmetries forming a larger order algebra via a map that is local from the un-free symmetry to the higher reducible one but non-local in the opposite direction. I will describe those mechanisms in detail and how they relate to the problem of communication through classical/quantum channels in a future article. What is important to notice here is that spacetime is not an "engineering" communication channel. Actually spacetime may be regarded as a strongly (higher topologically) entangled network in which only certain types of connections are possible, and those already encode much of the information that can be transmitted through them. For electromagnetism for example we are assuming a U(1) gauge invariance, but this needs to be that way in order to encode a massless particle with two states of helicity. Anyhow, avoiding such trivia, what is important is that spacetime, in its effective sense, does not act as an "engineering" channel, and indeed presents sensitivity to the correlations sent through it. But as said previously, even what we call classical information must have some quantum remnant for it to be able to describe interactions (aka gauge connections). To make this more precise we should consider the Reeh Schlieder theorem. This is a fundamental result of local quantum field theory that links local operations in a bounded and small region of spacetime, to resulting quantum states at arbitrary distance. Indeed, this results in long range correlations, and not in direct creation of specified states, therefore preserving unitarity. Its proof usually makes use of the edge of the wedge theorem that defines a connection between two "wedges" and one "edge" by means of holomorphic functions. This is seen as a source for entanglement in local quantum field theories, and it is a very strong statement that suggests that indeed spacetime (vacuum) is a correlation sensitive communication channel. 
In our case, the channel susceptible to correlations in the transmitted data is nothing but our stretching spacelike slice turned timelike inside the horizon. If we imagine the Planck scale fourier modes of our fields emerging from this stretching vacuum, we can consider that in this region we gain access to high energy small scale physics, and if the Planck scale is involved, we are free to assume that string modes are being taken into consideration. The mixing of scales is of course an interesting discussion all by itself, analysed in terms of T-duality in string theory and sigma models. What happened was for one channel of communication based on classical light pulse transmission to be blocked by means of a spacetime horizon, while another channel, this time one sensible to non-trivial higher topology correlations in vacuum, to be created in the form of the stretching spacelike slice around the horizon. This channel created entangled particles outside, slowly transferring all the information retained in the black hole, as the black hole evaporates. What is the difference between the two channels? The conventional channel requires a minimum of complexity as it is unaware of any correlations in the message it transmits, therefore considering every message to be "random noise". Its entropies are often calculated in the analysis of the black hole information paradox. The other channel is strongly "aware" of the correlations (a better word should probably be "sensitive") and therefore requires little information to be transmitted by the minimum complexity channel. In the extreme case, of a black hole that completely cuts off the causal (classical, but including the usual entanglement) channel, the correlation sensitive channel is all that remains, and hence all information needs to be transmitted at a high complexity rate. We call this channel "vacuum" or "wormhole" in common terms, but I suspect those terms are misleading and only suggestive for what is truly going on. Far from being something as dramatic as a wormhole, I suspect, at least at small scales, this channel to be a result of the intricacies of quantum communication that happens beyond the type of correlations permitted by the topological features considered previously. But then again, maybe a wormhole itself can be generalised to have such a quantum complexity interpretation and the two concepts will be dual. 
In this note I showed that the usual type of entanglement (as in non-separability of global information that has been constructed via causal signal transmission) is not sufficient to explain the black hole information paradox, and indeed, there could be a tower of higher types of topological entropies that could encode such information. Indeed, due to the existence of black holes, this "tower" can be re-arranged so that the first (classical, causal) terms become zero. At the same time, spacetime (vacuum) has been regarded as a communication channel sensitive to correlations within itself. These two concepts brought together shed a new light on our understanding of the black hole information paradox as well as on the nature of information.


\begin{thebibliography}{99}
\bibitem{1} Peter May, Some remarks on equivariant bundles and classifying spaces, Théorie de l’homotopie, Astérisque, no. 191 (1990), p. 15
\bibitem{2} A. Grothendieck , Revetements Etales et Groupe Fondamental - Seminaire de Geometrie Algebrique du Bois Marie 1960/61 (SGA 1) , LNM 224 Springer Heidelberg (1971)
\bibitem{3} Bao, N., Chatwin-Davies, A. Remmen, G.N. Entanglement of purification and multiboundary wormhole geometries. J. High Energ. Phys. 2019, 110 (2019).
\bibitem{4} P. Nguyen, T. Devakul, M.G. Halbasch, M.P. Zaletel and B. Swingle, Entanglement of purification: from spin chains to holography, JHEP 01 (2018) 098
\bibitem{5} T. Takayanagi and K. Umemoto, Entanglement of purification through holographic duality, Nature Phys. 14 (2018) 573
\bibitem{6} Akers, C., Engelhardt, N. Harlow, D. Simple holographic models of black hole evaporation. J. High Energ. Phys. 2020, 32 (2020).
\bibitem{7} A. Almheiri, R. Mahajan, J. Maldacena and Y. Zhao, The Page curve of Hawking radiation from semiclassical geometry, JHEP 03 (2020) 149
\bibitem{8} N. Engelhardt and A.C. Wall, Quantum extremal surfaces: holographic entanglement entropy beyond the classical regime, JHEP 01 (2015) 073
\bibitem{9} Gautason, F.F., Schneiderbauer, L., Sybesma, W. et al. Page curve for an evaporating black hole. J. High Energ. Phys. 2020, 91 (2020)
\bibitem{10} H. Liu and S. Vardhan, A dynamical mechanism for the Page curve from quantum chaos, arXiv:2002.05734
\bibitem{11} S. Mathur, D. Turton, The fuzzball nature of two-charge black hole microstates, Nuclear Phys. B, Vol 945, 114684 (2019)
\bibitem{12} S. Mathur, The information paradox: Conflicts and resolutions. Pramana - J Phys 79, 1059–1073 (2012).
\bibitem{13} Penington, G. Entanglement wedge reconstruction and the information paradox. J. High Energ. Phys. 2020, 2 (2020).
\bibitem{14} Xi Dong, Daniel Harlow, Aron C. Wall, Phys. Rev. Lett. 117, 021601 (2016)
\bibitem{15} A.C. Wall, Maximin surfaces, and the strong subadditivity of the covariant holographic entanglement entropy, Class. Quant. Grav. 31 (2014) 225007
\bibitem{16} Grant Salton, Brian Swingle, and Michael Walter, Phys. Rev. D 95, 105007 (2017)
\bibitem{17} C. Shannon, A mathematical theory of communication, The Bell System Technical Journal, Vol. 27, pp. 379–423, 623–656 (1948)
\bibitem{18} Andrei T. Patrascu, Phys. Rev. D 90, 045018 (2014)
\bibitem{19} V. A. Abakumova, I. Yu. Karataeva, S. L. Lyakhovich, Reducible Gauge symmetry versus unfree gauge symmetry in Hamiltonian formalisms, Nucl. Phys. B. Vol 973, 115577 (2021)
\bibitem{20} A. T. Patrascu, The Universal Coefficient Theorem and Quantum Field Theory, A Topological Guide for the Duality Seeker, ISBN: 978-3-319-46143-4 (2017)
\end{thebibliography}
\end{document}